\documentclass{aastex62}

\usepackage[T1]{fontenc}
\usepackage{ae,aecompl}
\usepackage{natbib} 

\submitjournal{ApJ}

\shorttitle{Episodic jets from black hole accretion disks}
\shortauthors{Shende, Subramanian \& Sachdeva}

\begin{document}

\title{Episodic Jets from Black Hole Accretion Disks}

\author{Mayur B. Shende}
\affil{Indian Institute of Science Education and Research \\
Dr. Homi Bhabha Road, Pashan, Pune 411008,
India}

\author{Prasad Subramanian}
\affiliation{Indian Institute of Science Education and Research \\
Dr. Homi Bhabha Road, Pashan, Pune 411008,
India}

\author{Nishtha Sachdeva}
\affiliation{Center for Space Environment Modeling
          Dept. of Climate and Space
          University of Michigan
          Ann Arbor MI, 48109
          USA
}


\begin{abstract}
Several active galactic nuclei and microquasars are observed to eject plasmoids that move at relativistic speeds. We envisage the plasmoids as pre-existing current carrying magnetic flux ropes that were initially anchored in the accretion disk-corona. The plasmoids are ejected outwards via a mechanism called the toroidal instability (TI). The TI, which was originally explored in the context of laboratory tokamak plasmas, has been very successful in explaining coronal mass ejections from the Sun. Our model predictions for plasmoid trajectories compare favorably with a representative set of multi-epoch observations of radio emitting knots from the radio galaxy 3C120, which were preceded by dips in Xray intensity.

\end{abstract}



\section{Introduction} 
\subsection{Relativistic jets}
Collimated, relativistic outflows from active galaxies harboring black holes continue to engage our attention since the discovery of the first optical jet in M87 \citep{curtis1918} and those of Cygnus A at radio wavelengths \citep{hargrave1974}. Sources in our galaxy have been shown to harbor miniature versions of such relativistic jets \citep{mirabel1994,fender1999,fender2004}. These remarkable observations have given rise to a host of theoretical models that address issues such as the initial formation and acceleration of such relativistic jets and the manner in which they remain collimated at large distances. Many of these models address the origin of steady jets and winds, ranging from Poynting flux-dominated jets (e.g., \citealt{blandford1977,lovelace1987}) to ones which address the mass loading of such jets (e.g., \citealt{Blandford1982,ustyugova1999}). While most of such steady state jet models appeal to electrons as their primary constituents, some of them (e.g., \citealt{contopoulos1995,sbk1999,le2004,le2005}) address the origin of hadrons in these jets.

\subsection{Discrete blobs}
In addition to a seemingly continuous background, several of these jets exhibit discrete blob-like structures moving at relativistic speeds e.g., the jet associated with the radio galaxy 3C 120 \citep{marscher2002, chatterjee2009, casadio2015,casadio2015b} and blobs in SS433 \citep{margon1979a,margon1979b} and GRS 1915+105 \citep{mirabel1994,fender1999,fender2004} and also in the gravitational wave source, GW170817 \citep{punsly2019}. Some authors envisage these blobs as concentrations of electrons accelerated by shocks travelling along the jet, while some think of them as discrete packets of plasma that are ejected from the central object. We adopt the latter scenario. Some models addressing the origin of such episodic emission appeal to instabilities or similar interruptions to the steady-state description of the accretion-ejection phenomenon (e.g., \citealt{stepanovs2014,mondal2018,chakrabarti2002,garain2019}). \cite{giannios2009} argue that spontaneous reconnection episodes in a Poynting flux dominated jet can produce small-scale jets (equivalently, small blobs of plasma) and \cite{mizuta2018} employ GRMHD simulations to suggest that the episodic variations in the accretion rate can produce intense Alfv\'enic pulses.



On the other hand, our Sun presents an instance of a steady state outflow (the solar wind) punctuated by occasional ejections of blobs (called coronal mass ejections; CMEs). CMEs from the solar corona have been extensively studied, both by way of observational data analysis \citep{gosling1974,hundhausen1984,bosman2012,gopalswamy2013} and models addressing their initiation mechanism \citep{lin2000,klimchuk2001,torok2004,zhang2005,kliem2006,isenberg2007,liu2016,cheng2017}.  \citet{yuan2009} have adapted the solar CME initiation model of \citet{lin2000} to explain the evolution and catastrophic ejection of a pre-existing flux rope plasmoid embedded in the corona of an accretion disk. The ejection process they appeal to is driven primarily by reconnection in the current sheet beneath the erupting plasmoid. In this paper, we explore the torus instability (TI) model \citep{kliem2006} as a means of launching plasmoids and compare our results with observations of the source 3C 120 \citep{marscher2002,chatterjee2009,casadio2015,casadio2015b}.

\section{The Torus Instability Model}

Before we describe the details of the TI model in the context of black hole accretion disk coronae, we note that the origin and dynamics of magnetic fields in accretion flows around black holes is still a subject of considerable investigation (e.g., \citealt{romanova1998,contopoulos2015,contopoulos2018}). However, the necessity of a magnetically structured corona in order to interpret X-ray observations has been recognized as far back as \citet{galeev1979}, and papers such as \citet{uzdensky2008} investigate this further. The corona can also be interpreted as the hot, bloated post-shock inner region of a two component advective flow (TCAF, \citealt{chakrabarti1995}).

The plasmoid is a curved, flux rope structure carrying a toroidal current embedded in the accretion disk corona (e.g., figure 1 of \citealt{yuan2009}).  It is well known from studies of laboratory tokamak plasmas that the magnetic hoop forces acting on such a structure cause them to expand outward. In the solar corona, the flux rope is nominally held in place by overlying (ambient) magnetic fields. If the overlying fields decrease fast enough with height, the flux rope is susceptible to an instability, causing it to erupt outwards \citep{kliem2006,torok2007,zuccarello2015,sachdeva2017}. Interestingly, it has been shown \citep{kliem2014,demoulin2010} that the TI model is equivalent to catastrophic loss of equilibrium models such as that of \citet{lin2000}.

We adapt the TI model of \citet{kliem2006} to our situation and prescribe the following equation of motion for a flux rope plasmoid:
\begin{equation}
\Gamma \rho_{m} \frac{d^{2}R}{dt^{2}} = \frac{I^{2}}{4 \pi^{2} b^{2} R^{2}} \bigg(L + \frac{\mu_{0} R}{2} \bigg) - \frac{I B_{\rm ext}(R)}{\pi b^2} - \frac{\rho_{m} c^2}{R_{\rm g} (R_* - 2)^2} \, \, \, . \label{eq:a}
\end{equation}

The quantity $\Gamma \equiv (1 - v^{2}/c^{2})^{-1/2}$ ($v$ is the velocity of the flux rope) is the bulk Lorentz factor of the flux rope, $\rho_m$ is the mass density inside it, $R$ and $b$ are the major and minor flux rope radii respectively, $I$ is the toroidal current and $B_{\rm ext}$ is the overlying poloidal magnetic field. Strictly speaking, the flux rope major radius ($R$) is equal to $\sqrt{z^{2}+r^{2}}$, where $r$ is the distance from the central object along the equatorial midplane of the accretion disk (essentially the radius in cylindrical coordinates) and $z$ is the height above the equatorial midplane. However, for practical purposes, we regard $R$ to be the height of the plasmoid above the equatorial midplane - an approximation that gets better with increasing height from the central object. The inductance of the flux rope, which is assumed to be slender ($R \gg b$), is given by $L = \mu_0 R [ln(8R/b) - 2 + l_i/2] $ \citep{landau}. For concreteness, we consider that the current density within the flux rope is uniform, which yields a value of $l_i = 1/2$ for the internal inductance per unit length. The quantity $R_{\rm g} \equiv GM/c^{2}$ is the gravitational radius and $R_{*}$ is the major radius in units of $R_{\rm g}$. From now on, all quantities with a subscript $*$ are in units of the gravitational radius. The first term on the right hand side of Eq~\ref{eq:a} describes the hoop force on the curved flux rope which causes it to expand outwards \citep{miyamoto1976,bateman1978}. The second term represents the competing Lorentz force arising from the current carried by the flux rope interacting with the external magnetic field $B_{\rm ext}$, which tends to hold it down. If the second term decreases fast enough with $R$, the flux rope is subject to the torus instability, which causes it to erupt \citep{kliem2006,bateman1978}. The third term on the right hand side of Eq~\ref{eq:a} represents the gravitational attraction of the black hole approximated by the pseudo-Newtonian potential appropriate to a Schwarzschild black hole \citep{paczynsky1980}. Our analysis does not include any terms arising from the gas pressure. If we were to include gas pressure effects inside the plasmoid, it would involve an additional term proportional to $(\overline{P} - P_{a})/B_{pa}^{2}$ inside the parantheses of the first term on the right hand side of Eq~\ref{eq:a} (e.g., \citealt{miyamoto1976,chen1996}). Here, $\overline{P}$ is the average gas pressure inside the plasmoid, $P_{a}$ is the ambient pressure outside the plasmoid and $B_{pa}$ denotes the poloidal magnetic field of the plasmoid evaluated at its outer boundary. However, it is widely understood that the ambient corona comprises a low-$\beta$ plasma, where the magnetic pressure dominates over the gas pressure (e.g., \citealt{liuqiao2016,rozanska2015} and also MRI simulations that address the vertical structure of the disk, such as \citealt{miller2000}). We will show in \S~3.3 below that the interior of the plasmoid is also a low-$\beta$ environment, as long as the proton temperature at the launching point is below $\approx 10^{11}$ K. Since both the plasmoid and the ambient environment are magnetically dominated, our neglect of thermal pressure terms in Eq~(\ref{eq:a}) is justified. We also note that our treatment does not address any possible departures from axisymmetry in the plasmoid trajectory. We will have occasion to comment on this later, in \S~3.2.

The total magnetic flux enclosed by the flux rope is 
\begin{equation}
\Psi = \Psi_I + \Psi_{\rm ext} = LI - 2 \pi \int_{0}^{R} B_{\rm ext}(r) rdr  \, , \label{eq:b}
\end{equation}
where the first term denotes the flux contained due to the magnetic fields inside the flux rope and the second that due to the external magnetic field enclosed by it. As in \citet{kliem2006}, we assume that the external field is unperturbed by the expanding flux rope and that it decreases with $R$ as 
\begin{equation}
B_{\rm ext}(R) = \hat{B} R^{-n}\, , \,\,\,\, n > 0 \, .
\label{eq:nn}
\end{equation}

The aspect ratio ($R/b$) of the flux rope is assumed to evolve as 
\begin{equation}
\frac{R}{b} = \frac{R_{0}}{b_{0}} \rho^p  \label{eq:d}
\end{equation}
where $\rho = R/R_{0}$ and $R_{0}/b_{0}$ is the aspect ratio at the equilibrium position of the plasmoid. The pre-eruption equilibrium position of the plasmoid ($R_{0}$) is determined by equating the right hand side of Eq~(\ref{eq:a}) to zero. For concreteness, we assume $p = 0$, which assumes that the aspect ratio of the plasmoid ($R/b$) remains constant as it evolves. The size $d = 2b$ of the plasmoid can be written as
\begin{equation}
d = 2b = 2b_{0} \rho^{1-p}	 \, ,			\label{eq:e}
\end{equation}
where $b_{0}$ denotes the flux rope minor radius at the pre-eruption equilibrium position. 

The statement of flux conservation (Eq~\ref{eq:b}) yields the following expression for the current enclosed by the flux rope at a given distance $R$ in terms of the current $I_{0}$ at the pre-eruption equilibrium position ($R_{0}$):
\begin{equation}
I(R) = \frac{a_0 R_{0} I_0}{a R} \bigg\{1 + \frac{2 \pi}{a_0 \mu_0 I_0} \frac{R_{0} B_{\rm eq}}{2-n} \bigg[ \bigg( \frac{R}{R_{0}} \bigg)^{2-n} - 1 \bigg] \bigg\}   \label{eq:c}
\end{equation}
where $B_{\rm eq} = B_{\rm ext}(R_0)$, $a = L/(\mu_0 R)$ and $a_0$ is value of $a$ at the pre-eruption equilibrium position $R_{0}$. When the expansion is self-similar, as we assume here ($p=0$), $a = a_{0}$.
The expression for current given in eq. (\ref{eq:c}) can be written in terms of the equilibrium current $I_{0}$ (determined by equating the right hand side of Eq~\ref{eq:a} to 0) as
\begin{equation}
I(R) = \frac{a_0 R_{0} I_0}{a R} \bigg\{ 1 + \frac{a_0 + 1/2}{a_0} \frac{1}{A (2-n)} \bigg[ \bigg( \frac{R}{R_{0}} \bigg)^{2-n} - 1 \bigg] \bigg\}        \label{eq:f}
\end{equation}
where
\begin{equation}
A = 1 + \sqrt{1 + \frac{\mu_0 c^2 b_{0}^2 \rho_{m0} (a_0 + 1/2)}{R_{0} R_{\rm g} B_{\rm eq}^2 (R_{0*} - 2)^2}} \, .  \label{eq:g}
\end{equation}

Inserting Eq. (\ref{eq:f}) in Eq. (\ref{eq:a}) yields

\begin{equation}
\frac{d^2 \rho}{d \tau^2} = \frac{a_0^2 (a + 1/2)}{4 a^2 (a_0 + 1/2)} A^2 Q^2 \rho^{-2} - \frac{a_0}{2 a}A Q \rho^{-n}  - \frac{T^2 c^2 R_{\rm g}}{\Gamma R_{0}^3} \frac{1}{\big(\rho - \frac{2}{R_{0*}}\big)^2}  
				\label{eq:h}
\end{equation}

where $\rho = R / R_{0}$ and $\tau = t/T$, with

\begin{equation}
T = \bigg( \frac{a_0 + 1/2}{4} \frac{b_{0}^2}{B_{\rm eq}^2 / \mu_0 \rho_{m0}} \bigg)^{1/2} = \frac{(a_0 + 1/2)^{1/2}}{2} \frac{b_{0}}{V_{\rm A_h}}   		 \label{eq:i}
\end{equation}
being the ``hybrid" Alfv\'en crossing time of the minor radius, $V_{\rm A_h} \equiv B_{\rm eq}/\sqrt{\mu_0 \rho_{m0}}$ the ``hybrid" Alfv\'en velocity and 
\begin{equation}
Q = 1 + \frac{(a_0 + 1/2)}{a_0} \frac{(\rho^{2-n} - 1)}{A (2-n)} \, .   			\label{eq:j}
\end{equation}

The bulk Lorentz factor of the plasmoid ($\Gamma$) can be expressed in terms of parameters $V_{\rm A_h}$ and $R_{0}/b_{0}$ as 
\begin{equation}
\Gamma = \bigg[ 1 - \bigg( \frac{R_{0}}{b_{0}} \bigg)^2 \frac{4 V_{\rm A_h}^2}{c^2 (a_0 + 1/2)} \bigg( \frac{d \rho}{d \tau} \bigg)^2 \bigg]^{-\frac{1}{2}}                     \label{eq:k}
\end{equation}
In non-relativistic limit ($v \ll c$, or equivalently, $\Gamma = 1$) and ignoring gravity, Eq. (\ref{eq:h}) reduces to the equation of motion quoted in \citet{kliem2006}.

\section{Results and Comparison With Observations}

\subsection{Velocity and acceleration profiles of a blob}
The flux rope parameters at the (pre-eruption) equilibrium position are:
\begin{eqnarray}
\nonumber
\rho(0) = 1  \\				
\rho'(0) = \frac{v_0 T}{R_{0}} = \beta_0 \frac{c T}{R_{0}}     \label{eq:l}
\end{eqnarray}
A small perturbation from the equilibrium position causes the flux rope to erupt outwards. The quantity $v_{0}$ denotes the launching velocity of the flux rope and $\beta_0 \equiv v_0 / c$. 

\begin{figure*}
\gridline{\fig{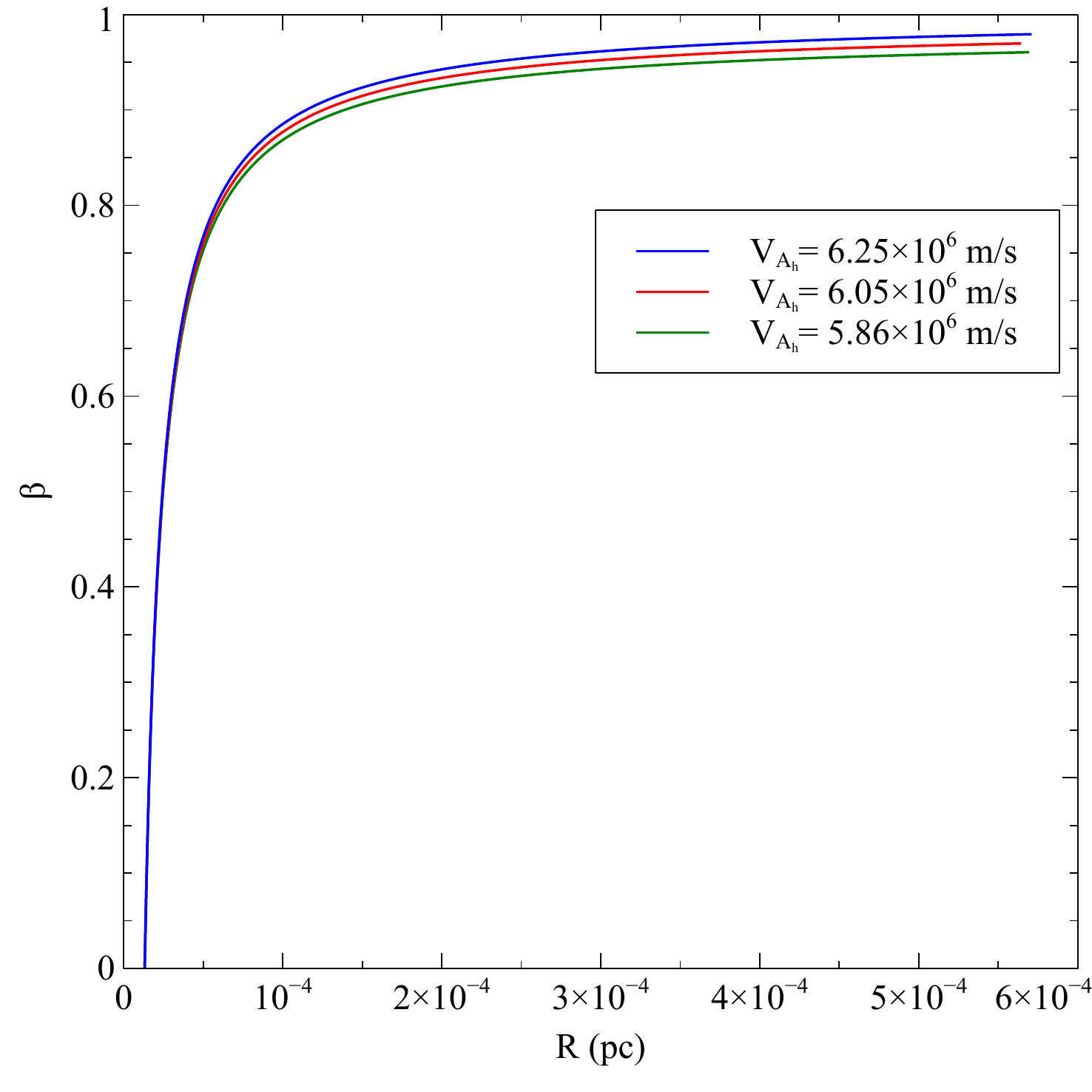}{0.46\textwidth}{(a)}
          \fig{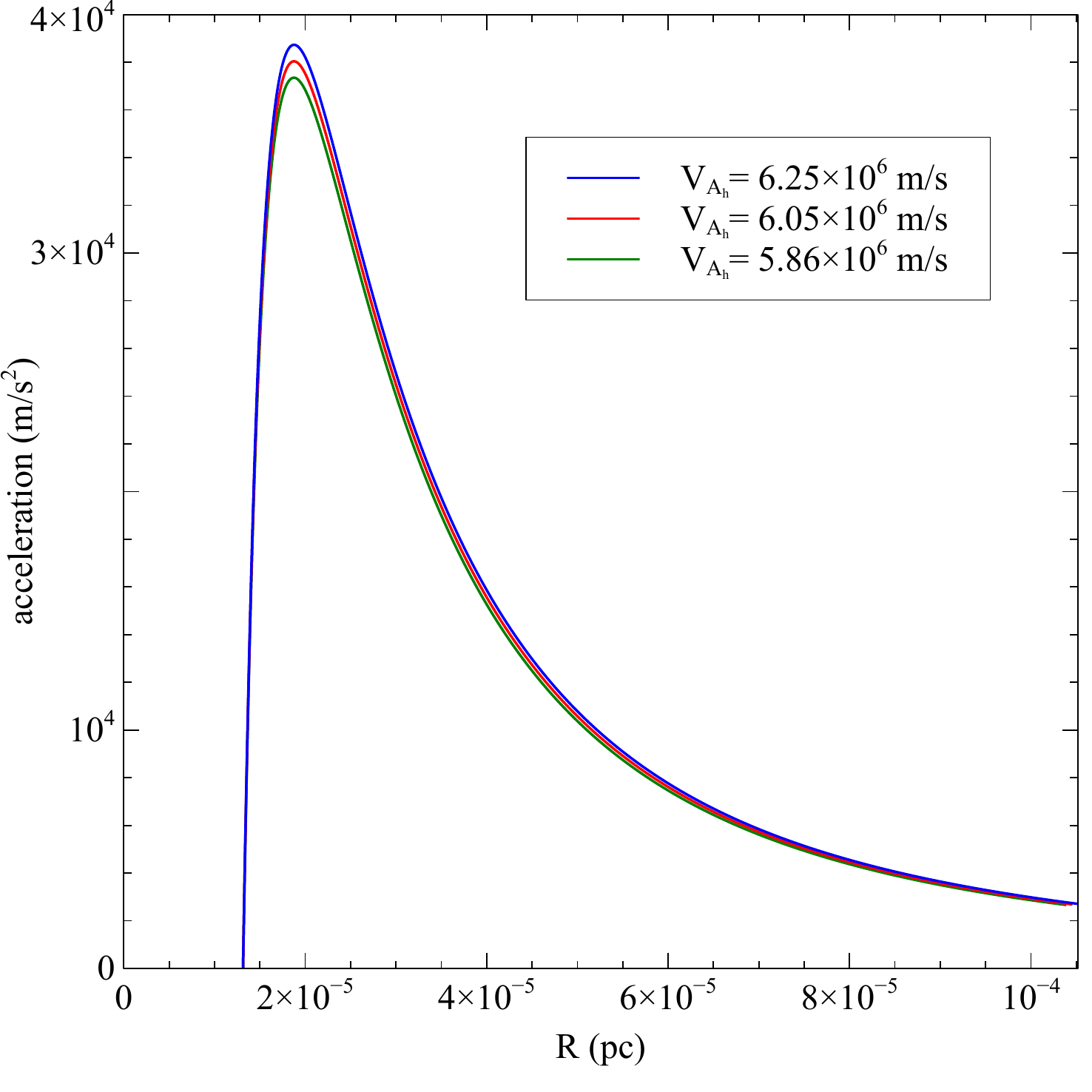}{0.46\textwidth}{(b)}
         }
 
\caption{(a) Velocity ($\beta \equiv v/c$) - distance profile and 
(b) Acceleration - distance profile of a blob for $M=5.5\times10^{7}M_{\odot}$, $R_{0}/b_{0}=10$, $R_{0}=5\,R_{\rm g}$, $n=4$, $p=0$ and $\beta_0 = 0.001$\label{fig:velacc}}
\end{figure*}

We solve the equation of motion (Eq~\ref{eq:a}) with the initial conditions specified in Eq~(\ref{eq:l}) and compare the results with multi-epoch observations of radio wavelength knots from 3C120. The parameters required to solve this equation are $M$, $n$, $R_{0}$, $R_{0}/b_{0}$, p, $V_{\rm A_h}$ and $\beta_0$. Before showing detailed comparisons with observations, we depict typical velocity and acceleration profiles of a representative blob in figures \ref{fig:velacc}(a) and \ref{fig:velacc}(b). The parameters used for these results are $p=0$ (self-similar expansion), $M=5.5\times10^{7}M_{\odot}$ (representative of 3C120 - \citealt{peterson2004}), $R_{0}/b_{0}=10$, $R_{0}=5 \,R_{\rm g}$, $n=4$ and $\beta_0 = 0.001$. The different linestyles use different values of Hybrid Alfv\'en velocities ($V_{\rm A_h}$). We have also examined the velocity and acceleration profiles by varying the other parameters. 
We note that increasing $V_{\rm A_h}$, $R_{0}/b_{0}$ and $\beta_0$ and decreasing $R_{0}$ results in lower values for the distance at which the blob acceleration reaches its peak. 
The broad conclusion from this exercise is that the acceleration profile peaks within  2 - 3 hours after launch, and the blob reaches its asymptotic speed soon thereafter. 
Since the ``core'' (which is the closest instance of a blob being imaged) for 3C120 is $\approx$ 0.22 pc from the central black hole \citep{chatterjee2009}, this implies that the multi-epoch radio observations can track the blobs' motion only well after they have attained their asymptotic speed.




\subsection{Comparison with multi-epoch observations of 3C120}
We compare the predictions of our model with imaging observations of blobs at radio frequencies from the nearby ($z=0.033$) radio galaxy 3C 120 \citep{marscher2002,chatterjee2009,casadio2015,casadio2015b}. While this source exhibits a morphology similar to FRI sources, it also has a blazar-like radio jet with blobs being ejected from the central source at regular intervals \citep{casadio2015,casadio2015b}. These multi-epoch observations at 15 and 43 GHz provide a detailed picture of blob position as a function of time, and \citet{gomez2000} suggest they are emitted at an angle $\sim$20$^\circ$ to the line of sight.
 \citet{chatterjee2009} show that pronounced dips in the X-ray luminosity precede the blob ejections by $\approx 66 \pm 51$ days. Since the X-ray emission likely originates in the accretion disk corona, this suggests a scenario where the coronal plasma is occasionally ``evacuated'' and ejected outwards as a blob, (much like a solar CME) which is later detected at radio frequencies. While \cite{marscher2002} were the first to suggest a connection between Xray dips and blob ejections in 3C120, \cite{nandi2001} also appeal to evacuation of the inner disk in trying to explain soft Xray dips in the galactic microquasar GRS 1915+105. In 3C120, the first observed instance of a blob (called the radio core) is at a distance of $\approx 0.22$ pc from the central object. Thereafter, the blob is imaged at successive epochs as it travels away from the central object. For the sake of concreteness, we concentrate on observations of blob E8 from \citet{casadio2015,casadio2015b} at 15 GHz. 

The apparent velocity of a blob ($\beta_{\rm app}$) as projected onto the plane of the sky is related to its ``true'' velocity ($\beta$) by
\begin{equation}
\beta_{\rm app}=\frac{\beta \sin \theta}{1-\beta \cos \theta} \, ,					\label{eq:m}
\end{equation}
where $\theta$ denotes the angle subtended by the blob's trajectory with the line of sight. For observations of 3C120, $\theta$ is estimated to be 20$^{\circ}$ \citep{gomez2000}.

\begin{figure*}
 \fig{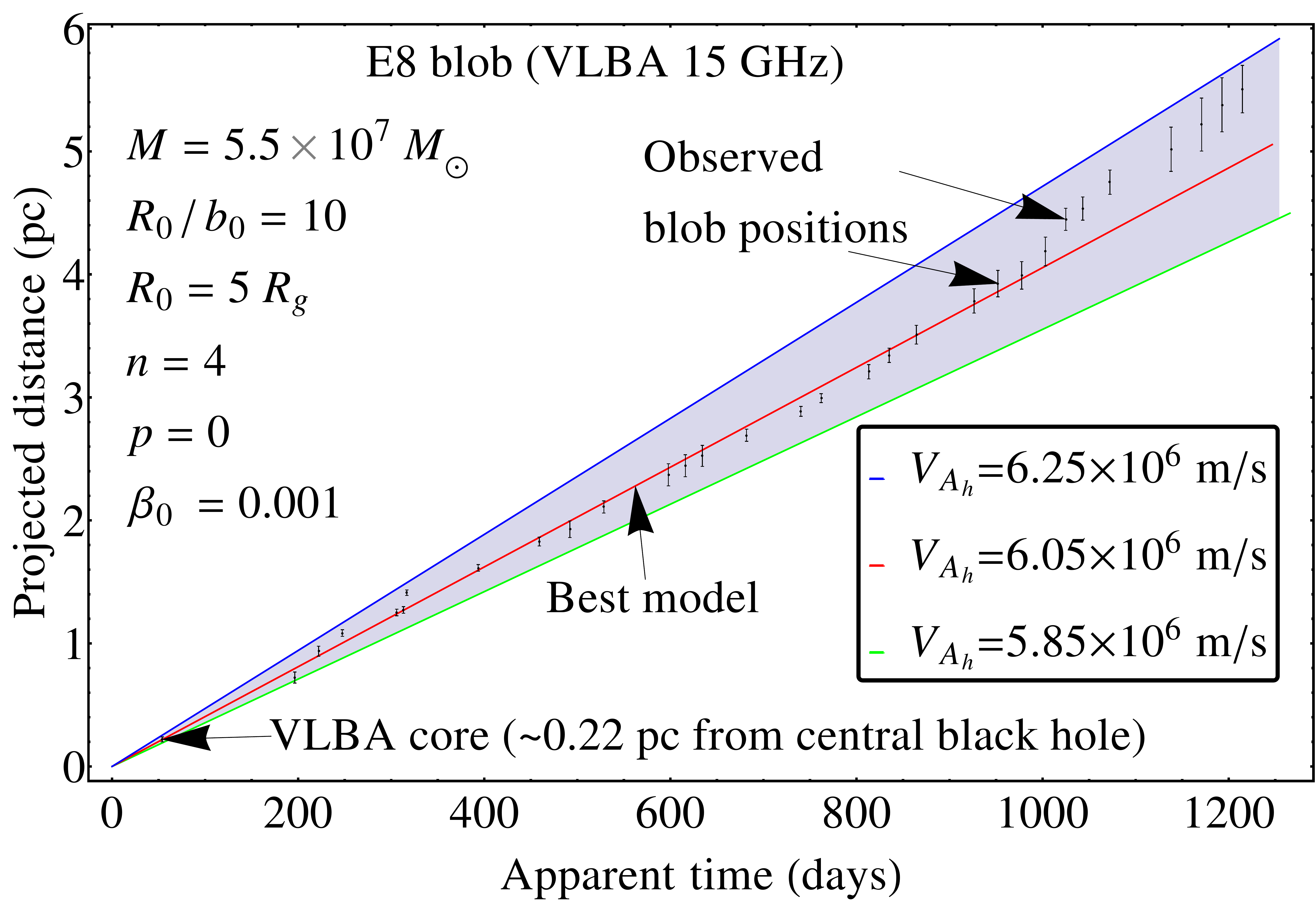}{0.60\textwidth}{}

 \caption{Height-time profile of a representative plasmoid for different values of $V_{\rm A_h}$ compared with observations of blob E8}
 \label{fig:htprofile}
\end{figure*}

The parameters yielding the best fit model to the observations of blob E8 are $M=5.5\times10^{7}M_{\odot}$, $R_{0}/b_{0}=10$, $R_{0}=5\, R_{\rm g}$, $n=4$, $V_{\rm A_h} = 6.05 \times 10^{6}$ ${\rm m/s}$ and $\beta_0 = 0.001$ which translates into launching Alfv\'enic Mach number $\mathcal{M}_{\rm A_h} \equiv v_{0}/V_{\rm A_h}= 0.05$. We note that $R_{0}=5\,R_{\rm g}$ can correspond to an equilibrium position that is on the equatorial accretion disk midplane and 5 gravitational radii from the black hole, or to one that is slightly closer to the black hole and slightly above the disk midplane. The classical version of the TI needs the ambient magnetic field ($B_{\rm ext}$) to decrease faster than a certain extent with height ($n > 1.5$, Eq~\ref{eq:nn}) in order for the instability to be operative \citep{kliem2006}. For large-scale magnetic fields in a black hole accretion disk-corona system, it is reasonable to expect that the ambient magnetic field decays faster than that of a large-scale dipole (for which $n=3$). Accretion disk dynamos relying on the magnetorotational instability predict that the large scale toroidal field has quadrupolar symmetry about the disk midplane \citep{brandenburg1995}. Recent GRMHD simulations \citep{mckinney2012} suggest a more complex scenario where the large-scale magnetic fields that develop depend upon the initially assumed magnetic field configuration. Some initial toroidal magnetic field configurations seem to generate patches of dipolar field. We adopt $n=4$ for our best fit fiducial model. The best fit fiducial model (depicted by the red line) together with the data for blob E8 are shown in figure \ref{fig:htprofile}. The blue and green curves show deviations from the best fit, which have values for $V_{\rm A_h}$ differing by +2.14 and -3.31\% (respectively) from that used for the best fit. The rest of the parameters are held fixed. The red line in figure~\ref{fig:htprofile}, which denotes the best fit to blob E8, seems to deviate somewhat from the observed positions at late times. In order to address this, we note that the plasmoid could possibly have non-axisymmetric motion (or wobble), which cannot be captured by the plane-of-sky observations; some of the deviation could be attributed to this. Furthermore, our model cannot account for non-axisymmetric motion. The blob corresponding to the best fit model takes $\approx 53$ days to travel the distance of 0.22 pc from the launching point. By comparison, the mean delay between the X-ray dip and the flaring of the radio core (which is interpreted as the time taken by the blob to travel the 0.22 pc between the black hole and the radio core) is $66 \pm 51$ days \citep{chatterjee2009}. The asymptotic superluminal velocity of the best fit model blob is 4.83$c$, while the observed value is $4.86c \pm 0.07c$ \citep{casadio2015,casadio2015b}.

While it is evident that our best fit model is in good agreement with the observations of blob E8 (figure~\ref{fig:htprofile}), it is worth checking how sensitive the model predictions are in response to changes in the parameters. We vary each parameter (while holding the rest of them fixed) to check how it affects the model prediction of the asymptotic blob velocity $\beta_{\rm app}$. The results of such a sensitivity study are shown in table \ref{tab:sensitivity}, and give an idea of the acceptable parameter space. We find that $\beta_{\rm app}$ is most sensitive to the hybrid Alfv\'en velocity $V_{A_h}$, equilibrium position $R_{0}$ and the initial aspect ratio $R_{0}/b_{0}$. It is only moderately sensitive to $n$, and quite insensitive to the launching velocity $\beta_0$. We note that the second term in Eq~(\ref{eq:k}) cannot exceed unity; this translates into upper limits on the allowed values for $V_{\rm A_h}$ and $R_{0}/b_{0}$.

We have also carried out a similar analysis for blob D11, which was observed at 43 GHz \citep{casadio2015,casadio2015b}. We find that the sensitivity of $\beta_{\rm app}$ to the changes in the parameters of best fit model for blob D11 is similar to that for E8.

\startlongtable
\begin{deluxetable}{c|c|c|c}
\tablecaption{An analysis of how the asymptotic $\beta_{app}$ for blob E8 changes with model parameters \label{tab:sensitivity}}
\tablehead{
\colhead{Parameter} & \colhead{} & \colhead{} & \colhead{$\%$ change} \\
\colhead{}          & \colhead{} & \colhead{} & \colhead{in $\beta_{app}$}
}

\startdata
    $V_{A_h}$ & Best fit    & 6050 $km/s$  \\
    \hline
              & $\%$ change & -3.31             & -12.63\\
              & $\%$ change & +2.14             & +16.15\\
    \hline
    $\beta_0$ & Best fit        & 0.001       \\
    \hline
                        & $\%$ change     & -99             & -0.02\\
                        & $\%$ change     & +900             & +0.1\\
    \hline
     $n$      & Best fit    & 4          \\
    \hline
              & $\%$ change & -10            & +1.41\\
              & $\%$ change & +10            & -0.93\\
    \hline
     $R_{0}$ & Best fit    & $5\,R_{\rm g}$  \\
    \hline
              & $\%$ change & -0.8           & +16.94\\
              & $\%$ change & +1             & -15.16\\
    \hline
     $R_{0}/b_{0}$ & Best fit    & 10 \\
    \hline
              & $\%$ change & -4             & -12.25\\
              & $\%$ change & +4             & +16.15\\
    \hline
    \hline
\enddata

\end{deluxetable}


\subsection{Mass loss rate due to blob ejection}

It is commonly understood that the observed radio frequency emission from the blobs is due to highly relativistic electrons gyrating in magnetic fields and emitting synchrotron emission. These energetic electrons are generally assumed to be confined within the blob by tangled magnetic fields. Our model has a definite prediction for the manner in which the plasmoid magnetic field varies with distance from the central object. Assuming that the plasmoid magnetic field is representative of the tangled field that confines the relativistic electrons enables us to estimate the mass ejection rate $\dot{M}_{\rm blob}$. 
For simplicity, we assume that the observed radio emission from the blob is at the ``critical'' synchrotron frequency ($\nu_{\rm c}$) and that the pitch angle is 90$^{\circ}$:
\begin{equation}
\nu_{\rm c} = \frac{3}{4 \pi} \frac{\gamma_{c}^2 e B_{\rm blob}}{m_{\rm e}}  \, , \label{eq:n}
\end{equation}

where $\gamma_{c}$ is the random Lorentz factor of the radio emitting electrons (as distinct from the quantity $\Gamma$ which denotes the bulk Lorentz factor of the blob). The quantity $B_{\rm blob}$ is the magnetic field carried by the blob and can be extracted from the first term on the RHS of Eq~(\ref{eq:a}):

\begin{equation}
B_{\rm blob} = \frac{a_0 (a + 1/2)}{2 a (a_0 + 1/2)} V_{A_h} \sqrt{\mu_0 n_0 (f m_p + m_e)} A Q \rho^{-2}    \label{eq:o}
\end{equation}

where $n_0$ is the number density of electrons at the (pre-eruption) equilibrium position of the plasmoid, $f$ is the ratio of protons to electrons in the blob, $m_p$ and $m_e$ are the masses of a proton and electron respectively and $A$ and $Q$ are given by Eqs. (\ref{eq:g}) and (\ref{eq:j}) respectively. 


Since the energetic electrons are confined inside the blob by the magnetic field, it follows that their Larmor radius ($r_{\rm L}$) is a fraction ($\alpha \leq 1$) of the comoving size ($d$) of the blob. In other words,
\begin{equation}
r_{\rm L} = \frac{\gamma_{c} m_{e} c}{e B_{\rm blob}} = \alpha d \, ,        \label{eq:p}
\end{equation}

where $d$ is comoving size of a blob given by Eq.(\ref{eq:e}). Using Eqs~\ref{eq:n}, \ref{eq:o} and \ref{eq:p}, the condition $\alpha \leq 1$ can be written as:
\begin{equation}
\alpha = \chi \bigg ( \frac{a}{a + 1/2} \bigg )^{3/2} \rho_{\rm ref}^{2+p} n_0^{-3/4}   \leq 1        \label{eq:q}
\end{equation}

where $$\chi = \frac{c}{b_0} \bigg ( \frac{m_e}{e} \bigg )^{3/2} \bigg ( \frac{\pi \nu_c}{3} \bigg )^{1/2} \bigg [ \frac{2 (a_0 + 1/2)}{a_0} \bigg ]^{3/2} (A V_{\rm A_h})^{-3/2} [\mu_0 (f m_p + m_e)]^{-3/4}$$ and $\rho_{\rm ref}$ represents the distance from the central object (in units of $R_{0}$) where the condition $\alpha \leq 1$ is imposed. Equivalently, the condition $\alpha \leq 1$ defines the following minimum value for $n_{0}$, the number density of electrons at the pre-eruption equilibrium position of the flux rope:

\begin{equation}
n_0 \geq \chi^{4/3} \bigg ( \frac{a}{a + 1/2} \bigg )^2 Q^{-2} \rho_{\rm ref}^{\frac{4 (2+p)}{3}}     \label{eq:r}
\end{equation}


The plasmoid magnetic field decreases with increasing distance from the central object; accordingly, the value for the confinement parameter $\alpha$ is expected to increase with distance. If, therefore, $\alpha = 1$ at the farthest observed position of a given blob, the radio emitting electrons are guaranteed to remain confined within it all the way from the pre-eruption equilibrium position until this point. 

The plasma-$\beta$ inside the blob/plasmoid is defined as 
\begin{equation}
\beta_{\rm blob} = \frac{\overline{P}}{B_{\rm blob}^2 / 2 \mu_0}           \label{eq:t}
\end{equation}
where $\overline{P}$ is the average gas pressure inside the blob.
For a magnetically dominated blob, $\beta_{\rm blob} < 1$ which translates into the following restriction on proton temperature: 
\begin{equation}
T_{p} < \frac{2 \,v_{\rm A_{blob}}^2 m_p}{k_{\rm B}}     \label{eq:u}
\end{equation}
where $k_{\rm B}$ is the Boltzmann constant and $v_{\rm A_{blob}} = B_{\rm blob} / \sqrt{\mu_0 m_p n}$ is the Alfv\'en velocity inside the blob. The quantity $n$ is the number density of protons which varies with the distance from the central black hole ($n = n_0$ at the launching point). We have also assumed here that $T_{p} \gg T_{e}$ ($T_{e}$ being the electron temperature) and that $f=1$ (i.e., there are equal numbers of electrons and protons). For the best fit model for blob E8, Eq~\ref{eq:u} predicts that the proton temperature $T_{p} \lesssim 10^{11}$K in order for $\beta_{\rm blob}$ to be $<1$. This is a very reasonable condition, and we therefore conclude that the plasmoid/blob is a low $\beta$ (i.e., magnetically dominated) plasma at launch.

For the representative blob E8 (observed at 15 GHz), the farthest observed position is $\approx 5.5$ pc \citep{casadio2015,casadio2015b}
Setting $\alpha = 1$ at a projected distance $R = 6$ pc from the central object (we use 6 pc instead of 5.5 pc for concreteness) yields $n_{0} = 4.88 \times 10^{15}\, {\rm m^{-3}}$.
The mass of the blob at the pre-eruption equilibrium position is related to $n_{0}$ via
\begin{equation}
M_{\rm blob} = 2 \pi^2 b_{0}^2 R_{0} n_0 (f m_p + m_e)  \, .   \label{eq:s} 
\end{equation}
The duration of the X-ray dips (5 to 120 days) quoted in \cite{chatterjee2009} are representative of the timescales over which the blobs are ejected from the accretion disk-corona system. Dividing the blob mass by this timescale allows us to estimate the blob mass ejection rate ($\dot{M}_{\rm blob}$). A useful quantity to compare the blob mass ejection rate with is the Eddington accretion rate $\dot{M}_{\rm E} \equiv L_{\rm E} / c^2$, where $L_{\rm E} \equiv 4 \pi G M c m_p / \sigma_{\rm T}$ is the Eddington luminosity and $\sigma_{\rm T}$ is the Thomson electron scattering cross section. Using $\nu_{c} = 15$ GHz, $f=1$ (equal numbers of electrons and protons), aspect ratio $R/b = 10$, $p=0$ (self-similar expansion) and $V_{\rm A_h} = 6050$ km/s (which corresponds to the ``best fit'' model for this blob; table~\ref{tab:sensitivity}), we find that $\dot{M}_{\rm blob} / \dot{M}_{\rm E}$ ranges from $\approx 10^{-6}$ (for an X-ray dip duration of 120 days) to $\approx 10^{-5}$ (for an X-ray dip duration of 5 days). In other words, the rate at which the blobs carry away mass from the accretion disk corona is quite small in comparison to the Eddington accretion rate. 
We note that requiring $\alpha = 1$ at 6 pc for our best-fit model implies an external magnetic field ($B_{\rm ext}$) of 194 G at the pre-eruption equilibrium position ($R_{0}$). On the other hand, magnetic fields as high as tens of kG near the central object are often invoked (e.g. \citealt{schwartz2019}) in order to satisfy observational constraints.  Requiring that $B_{\rm ext}$ at $R_{0}$ is larger by a factor of 100 translates to demanding $\alpha = 0.001$ at a distance of 6 pc from the central object. This in turn increases $n_{0}$ (Eq~\ref{eq:r}), the mass of the blob (Eq~\ref{eq:s}) and $\dot{M}_{\rm blob}$ by a factor of $10^{4}$, and $10^{-1} \gtrsim \dot{M}_{\rm blob} / \dot{M}_{\rm E} \gtrsim 10^{-2}$. The low values of $\dot{M}_{\rm blob}$ implies that most of the mass involved in the Xray dips is probably swallowed by the black hole, and only a small fraction is channeled into the outward directed plasmoids. This is broadly consistent with the expectations of similar scenarios in the galactic black hole source GRS1915+105 \citep{nandi2001}.

\section{Summary and Conclusions}
There is a fair amount of observational evidence suggesting that episodic blobs in AGN and microquasar jets are ejected from the accretion disk coronae surrounding the central black holes in these objects \citep{fender1999,belloni2001,fender2004,chatterjee2009,casadio2015,casadio2015b}. We consider the blob/plasmoid to be a curved, current carrying magnetic flux rope that is anchored in the accretion disk corona prior to eruption. This is a complementary approach to that of \citep{giannios2009,uzdensky2008} who consider the plasmoids to be spontaneously formed as a result of reconnection in the steady (Poynting flux dominated) jet. In our picture, the plasmoid is subject to two kinds of Lorentz forces - one is the co-called ``hoop force'', which arises in any curved flux tube carrying an axial current. Owing to the difference in magnetic pressures at the bottom and top of the bent flux tube \citep{mp1978}, this hoop force tends to push it outwards, away from the central object. The other kind of Lorentz force arises from the interaction between the current carried by the plasmoid and the magnetic fields external to it; this force tends to ``hold down'' the plasmoid and prevent it from erupting. If the magnitude of the external field decreases fast enough with distance from the central object, the plasmoid is subject to an ideal MHD instability (the TI), which causes it to erupt outwards like a whiplash. This effect has been applied with considerable success to explain the eruption of coronal mass ejections from the solar corona, and has held up to detailed comparison with observations \citep{sachdeva2017,gou2018}.

In this work, we have investigated the role of the TI in launching plasmoids from the accretion disk corona, and compared our model predictions with multi-epoch radio wavelength imaging observations of 3C120 \citep{chatterjee2009,casadio2015,casadio2015b}. This system exhibits dips in the Xray intensity that precede the ejection of plasmoids. Since the Xray emission is thought to originate in the corona, this lends support to a picture where parts of the coronal plasma are ejected in plasmoids (leading to Xray dips) and is subsequently imaged at radio wavelengths. We demonstrate that our model predictions for the time evolution of the ejected plasmoids agree well with the observed trajectories of the radio blobs in the 3C120 system. We also analyze the sensitivity of the model predictions to changes in the model parameters, by way of outlining a viable parameter space. It is commonly assumed that the highly relativistic electrons responsible for radio emission from the blobs are confined within it by tangled magnetic fields. We use this, together with other model assumptions and observational estimates of the Xray dip timescales to determine the mass loss rate due to plasmoid ejection. We find that this rate is typically a very small fraction of the Eddington accretion rate for 3C120. This is roughly similar to solar coronal mass ejections, which carry away a negligible fraction of the total mass contained in the solar corona. In summary, we have shown that the torus instability of flux rope plasmoids rooted in the accretion disk corona offers a convincing explanation for the observed episodic ejection of blobs from AGN and microquasars.

\section{Acknowledgements}

MBS acknowledges a PhD student fellowship from IISER, Pune. We acknowledge insightful comments from the anonymous referee that has helped us improve the paper.

\bibliography{ApJ11}

\begin{thebibliography}{}
\expandafter\ifx\csname natexlab\endcsname\relax\def\natexlab#1{#1}\fi
\providecommand{\url}[1]{\href{#1}{#1}}
\providecommand{\dodoi}[1]{doi:~\href{http://doi.org/#1}{\nolinkurl{#1}}}
\providecommand{\doeprint}[1]{\href{http://ascl.net/#1}{\nolinkurl{http://ascl.net/#1}}}
\providecommand{\doarXiv}[1]{\href{https://arxiv.org/abs/#1}{\nolinkurl{https://arxiv.org/abs/#1}}}

\bibitem[{{Bateman}(1978)}]{bateman1978}
{Bateman}, G. 1978, {MHD instabilities}

\bibitem[{{Belloni}(2001)}]{belloni2001}
{Belloni}, T. 2001, Astrophysics and Space Science Supplement, 276, 145

\bibitem[{{Blandford} \& {Payne}(1982)}]{Blandford1982}
{Blandford}, R.~D., \& {Payne}, D.~G. 1982, \mnras, 199, 883,
  \dodoi{10.1093/mnras/199.4.883}

\bibitem[{{Blandford} \& {Znajek}(1977)}]{blandford1977}
{Blandford}, R.~D., \& {Znajek}, R.~L. 1977, \mnras, 179, 433,
  \dodoi{10.1093/mnras/179.3.433}

\bibitem[{{Bosman} {et~al.}(2012){Bosman}, {Bothmer}, {Nistic{\`o}},
  {Vourlidas}, {Howard}, \& {Davies}}]{bosman2012}
{Bosman}, E., {Bothmer}, V., {Nistic{\`o}}, G., {et~al.} 2012, \solphys, 281,
  167, \dodoi{10.1007/s11207-012-0123-5}

\bibitem[{{Brandenburg} {et~al.}(1995){Brandenburg}, {Nordlund}, {Stein}, \&
  {Torkelsson}}]{brandenburg1995}
{Brandenburg}, A., {Nordlund}, A., {Stein}, R.~F., \& {Torkelsson}, U. 1995,
  \apj, 446, 741, \dodoi{10.1086/175831}

\bibitem[{{Casadio} {et~al.}(2015{\natexlab{a}}){Casadio}, {G{\'o}mez},
  {Grandi}, {Jorstad}, {Marscher}, {Lister}, {Kovalev}, {Savolainen}, \&
  {Pushkarev}}]{casadio2015}
{Casadio}, C., {G{\'o}mez}, J.~L., {Grandi}, P., {et~al.} 2015{\natexlab{a}},
  \apj, 808, 162, \dodoi{10.1088/0004-637X/808/2/162}

\bibitem[{{Casadio} {et~al.}(2015{\natexlab{b}}){Casadio}, {G{\'o}mez},
  {Grandi}, {Jorstad}, {Marscher}, {Lister}, {Kovalev}, {Savolainen}, \&
  {Pushkarev}}]{casadio2015b}
---. 2015{\natexlab{b}}, arXiv e-prints.
\newblock \doarXiv{arXiv:1505.03871}

\bibitem[{{Chakrabarti} \& {Titarchuk}(1995)}]{chakrabarti1995}
{Chakrabarti}, S., \& {Titarchuk}, L.~G. 1995, \apj, 455, 623,
  \dodoi{10.1086/176610}

\bibitem[{{Chakrabarti} {et~al.}(2002){Chakrabarti}, {Goldoni}, {Wiita},
  {Nandi}, \& {Das}}]{chakrabarti2002}
{Chakrabarti}, S.~K., {Goldoni}, P., {Wiita}, P.~J., {Nandi}, A., \& {Das}, S.
  2002, \apjl, 576, L45, \dodoi{10.1086/343104}

\bibitem[{{Chatterjee} {et~al.}(2009){Chatterjee}, {Marscher}, {Jorstad},
  {Olmstead}, {McHardy}, {Aller}, {Aller}, {L{\"a}hteenm{\"a}ki}, {Tornikoski},
  {Hovatta}, {Marshall}, {Miller}, {Ryle}, {Chicka}, {Benker}, {Bottorff},
  {Brokofsky}, {Campbell}, {Chonis}, {Gaskell}, {Gaynullina}, {Grankin},
  {Hedrick}, {Ibrahimov}, {Klimek}, {Kruse}, {Masatoshi}, {Miller}, {Pan},
  {Petersen}, {Peterson}, {Shen}, {Strel'nikov}, {Tao}, {Watkins}, \&
  {Wheeler}}]{chatterjee2009}
{Chatterjee}, R., {Marscher}, A.~P., {Jorstad}, S.~G., {et~al.} 2009, \apj,
  704, 1689, \dodoi{10.1088/0004-637X/704/2/1689}

\bibitem[{{Chen}(1996)}]{chen1996}
{Chen}, J. 1996, \jgr, 101, 27499, \dodoi{10.1029/96JA02644}

\bibitem[{{Cheng} {et~al.}(2017){Cheng}, {Guo}, \& {Ding}}]{cheng2017}
{Cheng}, X., {Guo}, Y., \& {Ding}, M. 2017, Science in China Earth Sciences,
  60, 1383, \dodoi{10.1007/s11430-017-9074-6}

\bibitem[{{Contopoulos} {et~al.}(2015){Contopoulos}, {Nathanail}, \&
  {Katsanikas}}]{contopoulos2015}
{Contopoulos}, I., {Nathanail}, A., \& {Katsanikas}, M. 2015, \apj, 805, 105,
  \dodoi{10.1088/0004-637X/805/2/105}

\bibitem[{{Contopoulos} {et~al.}(2018){Contopoulos}, {Nathanail}, {S{\c
  a}dowski}, {Kazanas}, \& {Narayan}}]{contopoulos2018}
{Contopoulos}, I., {Nathanail}, A., {S{\c a}dowski}, A., {Kazanas}, D., \&
  {Narayan}, R. 2018, \mnras, 473, 721, \dodoi{10.1093/mnras/stx2249}

\bibitem[{{Contopoulos} \& {Kazanas}(1995)}]{contopoulos1995}
{Contopoulos}, J., \& {Kazanas}, D. 1995, \apj, 441, 521,
  \dodoi{10.1086/175379}

\bibitem[{Curtis(1918)}]{curtis1918}
Curtis, H.~D. 1918, Pub. Lick Obs., 13, 31

\bibitem[{{D{\'e}moulin} \& {Aulanier}(2010)}]{demoulin2010}
{D{\'e}moulin}, P., \& {Aulanier}, G. 2010, \apj, 718, 1388,
  \dodoi{10.1088/0004-637X/718/2/1388}

\bibitem[{{Fender} \& {Belloni}(2004)}]{fender2004}
{Fender}, R., \& {Belloni}, T. 2004, \araa, 42, 317,
  \dodoi{10.1146/annurev.astro.42.053102.134031}

\bibitem[{{Fender} {et~al.}(1999){Fender}, {Garrington}, {McKay}, {Muxlow},
  {Pooley}, {Spencer}, {Stirling}, \& {Waltman}}]{fender1999}
{Fender}, R.~P., {Garrington}, S.~T., {McKay}, D.~J., {et~al.} 1999, \mnras,
  304, 865, \dodoi{10.1046/j.1365-8711.1999.02364.x}

\bibitem[{{Galeev} {et~al.}(1979){Galeev}, {Rosner}, \& {Vaiana}}]{galeev1979}
{Galeev}, A.~A., {Rosner}, R., \& {Vaiana}, G.~S. 1979, \apj, 229, 318,
  \dodoi{10.1086/156957}

\bibitem[{{Garain} {et~al.}(2019){Garain}, {Balsara}, {Chakrabarti}, \&
  {Kim}}]{garain2019}
{Garain}, S.~K., {Balsara}, D.~S., {Chakrabarti}, S.~K., \& {Kim}, J. 2019,
  private communication

\bibitem[{{Giannios} {et~al.}(2009){Giannios}, {Uzdensky}, \&
  {Begelman}}]{giannios2009}
{Giannios}, D., {Uzdensky}, D.~A., \& {Begelman}, M.~C. 2009, \mnras, 395, L29,
  \dodoi{10.1111/j.1745-3933.2009.00635.x}

\bibitem[{{G{\'o}mez} {et~al.}(2000){G{\'o}mez}, {Marscher}, {Alberdi},
  {Jorstad}, \& {Garc{\'{\i}}a-Mir{\'o}}}]{gomez2000}
{G{\'o}mez}, J.-L., {Marscher}, A.~P., {Alberdi}, A., {Jorstad}, S.~G., \&
  {Garc{\'{\i}}a-Mir{\'o}}, C. 2000, Science, 289, 2317,
  \dodoi{10.1126/science.289.5488.2317}

\bibitem[{{Gopalswamy}(2013)}]{gopalswamy2013}
{Gopalswamy}, N. 2013, in Astronomical Society of India Conference Series,
  Vol.~10, Astronomical Society of India Conference Series

\bibitem[{{Gosling} {et~al.}(1974){Gosling}, {Hildner}, {MacQueen}, {Munro},
  {Poland}, \& {Ross}}]{gosling1974}
{Gosling}, J.~T., {Hildner}, E., {MacQueen}, R.~M., {et~al.} 1974, \jgr, 79,
  4581, \dodoi{10.1029/JA079i031p04581}

\bibitem[{{Gou} {et~al.}(2018){Gou}, {Liu}, {Kliem}, {Wang}, \&
  {Veronig}}]{gou2018}
{Gou}, T., {Liu}, R., {Kliem}, B., {Wang}, Y., \& {Veronig}, A.~M. 2018, arXiv
  e-prints.
\newblock \doarXiv{1811.04707}

\bibitem[{{Hargrave} \& {Ryle}(1974)}]{hargrave1974}
{Hargrave}, P.~J., \& {Ryle}, M. 1974, \mnras, 166, 305,
  \dodoi{10.1093/mnras/166.2.305}

\bibitem[{{Hundhausen} {et~al.}(1984){Hundhausen}, {Sawyer}, {House}, {Illing},
  \& {Wagner}}]{hundhausen1984}
{Hundhausen}, A.~J., {Sawyer}, C.~B., {House}, L., {Illing}, R.~M.~E., \&
  {Wagner}, W.~J. 1984, \jgr, 89, 2639, \dodoi{10.1029/JA089iA05p02639}

\bibitem[{{Isenberg} \& {Forbes}(2007)}]{isenberg2007}
{Isenberg}, P.~A., \& {Forbes}, T.~G. 2007, \apj, 670, 1453,
  \dodoi{10.1086/522025}

\bibitem[{{Kliem} {et~al.}(2014){Kliem}, {Lin}, {Forbes}, {Priest}, \&
  {T{\"o}r{\"o}k}}]{kliem2014}
{Kliem}, B., {Lin}, J., {Forbes}, T.~G., {Priest}, E.~R., \& {T{\"o}r{\"o}k},
  T. 2014, \apj, 789, 46, \dodoi{10.1088/0004-637X/789/1/46}

\bibitem[{{Kliem} \& {T{\"o}r{\"o}k}(2006)}]{kliem2006}
{Kliem}, B., \& {T{\"o}r{\"o}k}, T. 2006, Physical Review Letters, 96, 255002,
  \dodoi{10.1103/PhysRevLett.96.255002}

\bibitem[{{Klimchuk}(2001)}]{klimchuk2001}
{Klimchuk}, J.~A. 2001, Washington DC American Geophysical Union Geophysical
  Monograph Series, 125, \dodoi{10.1029/GM125p0143}

\bibitem[{Landau {et~al.}(1984)Landau, Lifshitz, \& Pitaevskii}]{landau}
Landau, L.~D., Lifshitz, E.~M., \& Pitaevskii, L.~P. 1984, {Electrodynamics of
  continuous media; 2nd ed.}, Course of theoretical physics (Oxford:
  Butterworth).
\newblock \url{https://cds.cern.ch/record/712712}

\bibitem[{{Le} \& {Becker}(2004)}]{le2004}
{Le}, T., \& {Becker}, P.~A. 2004, \apjl, 617, L25, \dodoi{10.1086/427075}

\bibitem[{{Le} \& {Becker}(2005)}]{le2005}
---. 2005, \apj, 632, 476, \dodoi{10.1086/432927}

\bibitem[{{Lin} \& {Forbes}(2000)}]{lin2000}
{Lin}, J., \& {Forbes}, T.~G. 2000, \jgr, 105, 2375,
  \dodoi{10.1029/1999JA900477}

\bibitem[{{Liu} {et~al.}(2016{\natexlab{a}}){Liu}, {Qiao}, \&
  {Liu}}]{liuqiao2016}
{Liu}, J.~Y., {Qiao}, E.~L., \& {Liu}, B.~F. 2016{\natexlab{a}}, \apj, 833, 35,
  \dodoi{10.3847/1538-4357/833/1/35}

\bibitem[{{Liu} {et~al.}(2016{\natexlab{b}}){Liu}, {Kliem}, {Titov}, {Chen},
  {Wang}, {Wang}, {Liu}, {Xu}, \& {Wiegelmann}}]{liu2016}
{Liu}, R., {Kliem}, B., {Titov}, V.~S., {et~al.} 2016{\natexlab{b}}, \apj, 818,
  148, \dodoi{10.3847/0004-637X/818/2/148}

\bibitem[{{Lovelace} {et~al.}(1987){Lovelace}, {Wang}, \&
  {Sulkanen}}]{lovelace1987}
{Lovelace}, R.~V.~E., {Wang}, J.~C.~L., \& {Sulkanen}, M.~E. 1987, \apj, 315,
  504, \dodoi{10.1086/165156}

\bibitem[{{Margon} {et~al.}(1979{\natexlab{a}}){Margon}, {Ford}, {Grandi}, \&
  {Stone}}]{margon1979a}
{Margon}, B., {Ford}, H.~C., {Grandi}, S.~A., \& {Stone}, R.~P.~S.
  1979{\natexlab{a}}, \apjl, 233, L63, \dodoi{10.1086/183077}

\bibitem[{{Margon} {et~al.}(1979{\natexlab{b}}){Margon}, {Ford}, {Katz},
  {Kwitter}, {Ulrich}, {Stone}, \& {Klemola}}]{margon1979b}
{Margon}, B., {Ford}, H.~C., {Katz}, J.~I., {et~al.} 1979{\natexlab{b}}, \apjl,
  230, L41, \dodoi{10.1086/182958}

\bibitem[{{Marscher} {et~al.}(2002){Marscher}, {Jorstad}, {G{\'o}mez}, {Aller},
  {Ter{\"a}sranta}, {Lister}, \& {Stirling}}]{marscher2002}
{Marscher}, A.~P., {Jorstad}, S.~G., {G{\'o}mez}, J.-L., {et~al.} 2002, \nat,
  417, 625, \dodoi{10.1038/nature00772}

\bibitem[{{McKinney} {et~al.}(2012){McKinney}, {Tchekhovskoy}, \&
  {Blandford}}]{mckinney2012}
{McKinney}, J.~C., {Tchekhovskoy}, A., \& {Blandford}, R.~D. 2012, \mnras, 423,
  3083, \dodoi{10.1111/j.1365-2966.2012.21074.x}

\bibitem[{{Miller} \& {Stone}(2000)}]{miller2000}
{Miller}, K.~A., \& {Stone}, J.~M. 2000, \apj, 534, 398, \dodoi{10.1086/308736}

\bibitem[{{Mirabel} \& {Rodr{\'{\i}}guez}(1994)}]{mirabel1994}
{Mirabel}, I.~F., \& {Rodr{\'{\i}}guez}, L.~F. 1994, \nat, 371, 46,
  \dodoi{10.1038/371046a0}

\bibitem[{{Miyamoto}(1980)}]{miyamoto1976}
{Miyamoto}, K. 1980, {Plasma physics for nuclear fusion}

\bibitem[{{Mizuta} {et~al.}(2018){Mizuta}, {Ebisuzaki}, {Tajima}, \&
  {Nagataki}}]{mizuta2018}
{Mizuta}, A., {Ebisuzaki}, T., {Tajima}, T., \& {Nagataki}, S. 2018, \mnras,
  479, 2534, \dodoi{10.1093/mnras/sty1453}

\bibitem[{{Mondal} \& {Mukhopadhyay}(2018)}]{mondal2018}
{Mondal}, T., \& {Mukhopadhyay}, B. 2018, \mnras, 476, 2396,
  \dodoi{10.1093/mnras/sty332}

\bibitem[{{Mouschovias} \& {Poland}(1978)}]{mp1978}
{Mouschovias}, T.~C., \& {Poland}, A.~I. 1978, \apj, 220, 675,
  \dodoi{10.1086/155951}

\bibitem[{{Nandi} {et~al.}(2001){Nandi}, {Chakrabarti}, {Vadawale}, \&
  {Rao}}]{nandi2001}
{Nandi}, A., {Chakrabarti}, S.~K., {Vadawale}, S.~V., \& {Rao}, A.~R. 2001,
  \aap, 380, 245, \dodoi{10.1051/0004-6361:20011444}

\bibitem[{{Paczy{\'n}sky} \& {Wiita}(1980)}]{paczynsky1980}
{Paczy{\'n}sky}, B., \& {Wiita}, P.~J. 1980, \aap, 88, 23

\bibitem[{{Peterson} {et~al.}(2004){Peterson}, {Ferrarese}, {Gilbert}, {Kaspi},
  {Malkan}, {Maoz}, {Merritt}, {Netzer}, {Onken}, {Pogge}, {Vestergaard}, \&
  {Wandel}}]{peterson2004}
{Peterson}, B.~M., {Ferrarese}, L., {Gilbert}, K.~M., {et~al.} 2004, \apj, 613,
  682, \dodoi{10.1086/423269}

\bibitem[{{Punsly}(2019)}]{punsly2019}
{Punsly}, B. 2019, arXiv e-prints.
\newblock \doarXiv{1901.08224}

\bibitem[{{Romanova} {et~al.}(1998){Romanova}, {Ustyugova}, {Koldoba},
  {Chechetkin}, \& {Lovelace}}]{romanova1998}
{Romanova}, M.~M., {Ustyugova}, G.~V., {Koldoba}, A.~V., {Chechetkin}, V.~M.,
  \& {Lovelace}, R.~V.~E. 1998, \apj, 500, 703, \dodoi{10.1086/305760}

\bibitem[{{R{\'o}{\.z}a{\'n}ska} {et~al.}(2015){R{\'o}{\.z}a{\'n}ska},
  {Malzac}, {Belmont}, {Czerny}, \& {Petrucci}}]{rozanska2015}
{R{\'o}{\.z}a{\'n}ska}, A., {Malzac}, J., {Belmont}, R., {Czerny}, B., \&
  {Petrucci}, P.-O. 2015, \aap, 580, A77, \dodoi{10.1051/0004-6361/201526288}

\bibitem[{{Sachdeva} {et~al.}(2017){Sachdeva}, {Subramanian}, {Vourlidas}, \&
  {Bothmer}}]{sachdeva2017}
{Sachdeva}, N., {Subramanian}, P., {Vourlidas}, A., \& {Bothmer}, V. 2017,
  \solphys, 292, 118, \dodoi{10.1007/s11207-017-1137-9}

\bibitem[{{Schwartz}(2019)}]{schwartz2019}
{Schwartz}, D.~A. 2019, arXiv e-prints.
\newblock \doarXiv{1901.00462}

\bibitem[{{Stepanovs} {et~al.}(2014){Stepanovs}, {Fendt}, \&
  {Sheikhnezami}}]{stepanovs2014}
{Stepanovs}, D., {Fendt}, C., \& {Sheikhnezami}, S. 2014, \apj, 796, 29,
  \dodoi{10.1088/0004-637X/796/1/29}

\bibitem[{{Subramanian} {et~al.}(1999){Subramanian}, {Becker}, \&
  {Kazanas}}]{sbk1999}
{Subramanian}, P., {Becker}, P.~A., \& {Kazanas}, D. 1999, \apj, 523, 203,
  \dodoi{10.1086/307703}

\bibitem[{{T{\"o}r{\"o}k} \& {Kliem}(2004)}]{torok2004}
{T{\"o}r{\"o}k}, T., \& {Kliem}, B. 2004, in ESA Special Publication, Vol. 575,
  SOHO 15 Coronal Heating, ed. R.~W. {Walsh}, J.~{Ireland}, D.~{Danesy}, \&
  B.~{Fleck}, 56

\bibitem[{{T{\"o}r{\"o}k} \& {Kliem}(2007)}]{torok2007}
{T{\"o}r{\"o}k}, T., \& {Kliem}, B. 2007, Astronomische Nachrichten, 328, 743,
  \dodoi{10.1002/asna.200710795}

\bibitem[{{Ustyugova} {et~al.}(1999){Ustyugova}, {Koldoba}, {Romanova},
  {Chechetkin}, \& {Lovelace}}]{ustyugova1999}
{Ustyugova}, G.~V., {Koldoba}, A.~V., {Romanova}, M.~M., {Chechetkin}, V.~M.,
  \& {Lovelace}, R.~V.~E. 1999, \apj, 516, 221, \dodoi{10.1086/307093}

\bibitem[{{Uzdensky} \& {Goodman}(2008)}]{uzdensky2008}
{Uzdensky}, D.~A., \& {Goodman}, J. 2008, \apj, 682, 608,
  \dodoi{10.1086/588812}

\bibitem[{{Yuan} {et~al.}(2009){Yuan}, {Lin}, {Wu}, \& {Ho}}]{yuan2009}
{Yuan}, F., {Lin}, J., {Wu}, K., \& {Ho}, L.~C. 2009, \mnras, 395, 2183,
  \dodoi{10.1111/j.1365-2966.2009.14673.x}

\bibitem[{{Zhang} \& {Low}(2005)}]{zhang2005}
{Zhang}, M., \& {Low}, B.~C. 2005, \araa, 43, 103,
  \dodoi{10.1146/annurev.astro.43.072103.150602}

\bibitem[{{Zuccarello} {et~al.}(2015){Zuccarello}, {Aulanier}, \&
  {Gilchrist}}]{zuccarello2015}
{Zuccarello}, F.~P., {Aulanier}, G., \& {Gilchrist}, S.~A. 2015, \apj, 814,
  126, \dodoi{10.1088/0004-637X/814/2/126}

\end{thebibliography}



\end{document}